\documentclass[aps,eqsecnum,showpacs,prb,preprint,superscriptaddress]{revtex4}
\usepackage[pdftex]{graphicx}
\usepackage[english]{babel} 
\usepackage{amsmath}
\usepackage[latin1]{inputenc}
\usepackage{subfigure}

\begin{document}
\title{STM Simulation of Molecules on Ultrathin Insulating Overlayers
  using Tight-Binding: Au-Pentacene on NaCl bilayer on Cu}
\author{Antti Korventausta}
\affiliation{
Department of Physics, Tampere University Of Technology, Finland}
\author{Sami Paavilainen}
\affiliation{
Department of Physics, Tampere University Of Technology, Finland}
\author{Eeva Niemi}
\affiliation{
  CEMES-CNRS, 29 rue J. Marvig, P.O. Box 4347, F-31055 Toulouse cedex, France}
\author{Jouko Nieminen}
\affiliation{
Department of Physics, Tampere University Of Technology, Finland}

\date{\today}

\begin{abstract}
  We present a fast and efficient tight-binding (TB) method for
  simulating scanning tunneling microscopy (STM) imaging of adsorbate
  molecules on ultrathin insulating films.  Due to the electronic
  decoupling of the molecule from the metal surface caused by the
  presence of the insulating overlayer, the STM images of the frontier
  molecular orbitals can be simulated using a very efficient scheme,
  which also enables the analysis of phase shifts in the STM current.
  Au-pentacene complex adsorbed on a NaCl bilayer on Cu substrate
  provides an intricate model system, which has been previously
  studied both experimentally and theoretically. Our calculations
  indicate that the complicated shape of the molecular orbitals may
  cause multivalued constant current surfaces --- leading to
  ambiguity of the STM image.  The results obtained using the TB
  method are found to be consistent with both DFT calculations and
  experimental data.
\end{abstract}
\maketitle

\section{Introduction}

The continuous miniaturization of electronics to obtain more efficient
and smaller components is approaching the classical border beyond
which components consisting of single molecules are one possible
solution. In the development of such molecular components, an
excellent platform is provided by ultrathin insulating overlayers on
metal surfaces combined with scanning tunneling microscopy
(STM). Adsorbates on such films are electronically decoupled from the
metal substrate, which can be used, e.g., in controlling the charge
state of the adsorbate\cite{repp2} or studying atoms and molecules
without the influence of the metal background\cite{repp,paavi1}.

For complicated systems theoretical calculations play a critical role
in the interpretation of the experiments. However, many of the
currently studied organic molecules suitable for molecular electronic
devices are of a scale beyond a feasible system size for
\textit{ab-initio} calculations including the substrate. Semiempirical
tight-binding (TB) based methods provide an alternative solution to
this problem, allowing for simulation of larger systems with
reasonable computational resources. Additionally, TB methods provide
different kind of tools to analyze both the adsorption and the STM
images.

In this work we study a system consisting of a pentacene molecule
bonded with a gold atom on a NaCl overlayer previously studied both
experimentally and with \textit{ab-initio} methods.\cite{paavi1} Due
to its complicate electronic and geometric structure, this system
provides an excellent test ground for developing TB based simulations
of STM on molecules on insulating overlayers. The system exhibits
clearly recognizable properties caused by the interaction of the
pentacene and the gold atom, and by the adsorption of the complex on
the insulating overlayer, making it ideal for this work.

We use a purpose-built method for simulating STM in a TB basis
\cite{nieminen}, which has previously been applied to studying
superconducting materials and organic molecules on metal surfaces
\cite{niemi1}. This method allows simulations with functionalized
tips, analysis of tunneling channels, and studying of the molecular
orbitals of adsorbed molecules. Here we develop the method further by
introducing an even faster method for simulating STM of systems where
a certain molecular orbital dominates the STM image. Provided that
there is only a single molecular energy level within the bias voltage,
the STM current as a function of position is approximately
proportional to the hopping integral between the STM tip and the
frontier molecular orbital. This also allows studying of extremely
large systems, or rapidly exploring different molecular
configurations. The method developed in this work also makes it
possible to study the phase information of the STM current, which could
be utilized, e.g. in the simulation of isospectral molecules
\cite{Moon2008}.

This paper is organized as follows: Sec. \ref{sec:theory} explains the
theory behind the used calculation methods, Sec. \ref{sec:model}
details the model used for the Au-pentacene calculations,
Sec. \ref{sec:results} presents the obtained results, and
Sec. \ref{sec:summary} discusses the results and their
implications. In Appendix \ref{sec:appendix} several different forms
of calculating tunneling current are discussed.

\section{Theory}
\label{sec:theory}

We use two different approaches, one based on direct diagonalization of the
Hamiltonian and the other on Green's function formalism, to calculate
the electronic structure in a tight-binding basis. The former approach
is used to visualize the molecular orbitals and simulate their
behaviour in bond formation and adsorption. The latter method allows
detailed analysis of the electron transport properties and is used in
simulating STM images. 

In both cases the Hamiltonian is presented in atomic orbital
basis. The on-site terms (energies) are parameters fitted to reproduce
the density of states compared to density functional theory (DFT)
calculations, or taken from literature. The off-diagonal elements are
calculated using modified Slater-Koster hopping integrals described in
Sec. \ref{sec:model}.

In the first method the Hamiltonian is diagonalized using the secular
equations $\mathbf{H}\mathbf{c}^i=E_i\mathbf{c}^i$ in the H\"uckel
approximation where overlap between different atomic orbitals is
neglected. The obtained eigenvectors $\mathbf{c}^i$ include weight
coefficients $c_\alpha^i=\langle \alpha | i \rangle $ of individual
atomic orbitals $\alpha$ to an eigenstate $i$ of the system with
eigenenergy $E_i$. The weight coefficients allow real space projections
of the molecular orbitals by using the spatial dependence of Slater
type orbitals. This makes it also possible to simulate STM with the
Tersoff-Hamann approach\cite{teha} (TH) in which surfaces of the
constant local density of the sample states at the position of the STM
tip are identified with the topographic STM images.

In addition, the contribution of a molecular orbital $\mu$ of
a free molecule to an electronic state $i$ of a more complicated
system can be obtained by solving the secular equations separately for
the isolated molecule (non-interacting system) and for the interacting
system consisting of the molecule and other atoms. Using the
eigenvectors $\mathbf{c}^{m}$ and $\mathbf{c}^i$ calculated in the
same basis set $\alpha$ we obtain the weight of molecular orbital $\mu$
on the eigenstate $i$:
\begin{equation}
  c^i_\mu = \langle \mu | i  \rangle = \sum_\alpha \langle \mu | \alpha  \rangle 
  \langle \alpha | i \rangle =  \sum_\alpha c_\alpha^{\mu*}
  c^i_\alpha.
  \label{coeffeq}
\end{equation}
$\mathbf{c}^i_\mu$ is actually the eigenvector of Hamiltonian
represented in a mixed basis set consisting of molecular orbitals of
the free molecule and the atomic orbitals of the other atoms of the
system not involved in the free molecule.

This formalism can be used both to study the changes in molecular
states in the formation of chemical bonds, as well as to locate the
molecular states of a free molecule from an adsorbed system. This also
gives information on the interaction of the molecular orbitals with
the surface states. Since diagonalization of the Hamiltonian is very
fast, this method allows studying of very large systems.

In simulating STM images we use a Green's functions approach to
calculate the electronic structure of the sample, described in detail
in Ref. \onlinecite{nieminen}. The advantage of this approach is
obtaining the off-diagonal terms necessary to calculate currents
within Todorov-Pendry scheme (TP)\cite{todorov,pendry}.  This enables
detailed analysis of the STM image; the current can be decomposed to
the individual contributions of particularly important atomic orbitals
which form tunneling channels\cite{niemi1}. As shown in
Eq.\ref{currentV} in Appendix \ref{sec:appendix}, the current flowing
from the STM tip to the substrate can be written as
\begin{eqnarray}
  \label{eq:TPcurrent}
  I(V,r_t) &=& \frac{2e\eta}{\hbar} \int_0^{eV} j(E,r_t) dE\\
  j(E,r_t) 
  &=&  \,\mathrm{Tr}[\rho_{t't}^{0}(E) V_{ts'}(r_t) \nonumber
  G_{s'f}^-(E) G_{fs}^+(E) V_{st'}(r_t)]
\end{eqnarray}
where $r_t$ denotes the tip position and $V$ the bias voltage. Here
$j(E,r_t)$ is the differential conductance in units of
$\frac{2e\eta}{\hbar}$ corresponding to bias $E/e$. Subindices $f$,
$t$ and $s$ refer to final, tip and sample states,
respectively. $G_{\alpha\beta}^{0-}$ and $G_{\alpha\beta}^{0+}$ are
the retarded and advanced Green's function matrices, respectively, and
$2\eta$ is the full width at half maximum of the broadened delta
function used in calculating the Green's functions.
$V_{ts}(r_t)=\langle t(r_t) | H | s \rangle$ is the hopping integral
from tip to sample state $s$. The density matrix of the tip
$\rho_{t't}^{0}(E)$ is calculated separately from the sample.

On ultrathin insulating films, the STM experiments on adsorbates can
produce images whose shape is dominated by a single (frontier) orbital
of the adsorbate molecule\cite{repp}. In the following we show that
for this kind of system, STM images can be simulated by {\it hopping
  maps} in which the hopping integral from tip to the dominating state
of the system is plotted as a function of tip position. The idea lies
in the decoupling of electronic states of molecule and the underlying
metal substrate, due to which there is a single eigenstate of the
system which resembles one of the original molecular orbitals of the
adsorbate. Thus, there is a subset of sample states (refered to as
$\mu'$) which correspond to the orbitals of a free molecule (refered
to as $\mu$).

First, the tip is approximated by a single electron state so that its
density matrix element $\rho_{tt}^0$ can be taken outside the
calculation of matrix trace in Eq. \ref{eq:TPcurrent}. In addition,
the trace can be written as a double sum
\begin{equation}
  j(E,r_t) = \rho_{tt}^0(E) \sum_{s} \sum_{f} |V_{ts}(r_t) G_{sf}^-(E)|^2.
\end{equation}

Next we diagonalize the Hamiltonian of the sample using the secular
equation and rewrite the summation keeping in mind that the $V_{ts}$
and $c_\alpha^i$ are chosen to be real:
\begin{equation}
  j(E,r_t) = \rho_{tt}^0(E) \sum_{i} \sum_{f} |V_{ti}(E)|^2|G_{if}^-(E)|^2.
\end{equation}
If in the energy range 
$\left[E_\textrm{F},E_\textrm{F}+eV\right]$ defined by the bias voltage $V$ there is only a
single eigenstate $\nu'$ of the subset $\mu'$ of the sample states 
then we can assume that for all the other states
\begin{equation}
  |V_{ti}G_{if}(E)| \ll |V_{t\nu'}G_{\nu'f}(E)|, 
  ~\textrm{for all}~ i \neq \nu' \land E \in \left[ E_\textrm{F}, E_\textrm{F}+eV  \right].
\end{equation}
The validity of this assumption lies on the nature of the sample
states: either the Green's function is very small since the state is
too far away from the Fermi level (valid for the subset $\mu'$ and the
states corresponding to the atoms of the insulating film), or the
exponentially decaying hopping integral is negligible due to long
inter-atomic distance (valid for the metal states). Now the tunneling
current of Eq. \ref{eq:TPcurrent} can be simplified to
\begin{equation}
  I(V,r_t) \approx \frac{2e\eta}{\hbar} |V_{t\nu'}(r_t)|^2
  \int_0^{eV} \rho_{tt}^0(E) \sum_{f} |G_{\nu'f}^-(E)|^2,
\end{equation}
where the only dependence on the position of the STM tip, $r_t$, is
outside of the integral which is constant for a certain bias and
molecular orbital. Thus, for a bias voltage corresponding to a suitable 
energy range, the spatial dependence of the STM current can be written as:
\begin{equation}
  I(r_t) \propto |V_{t\nu'}(r_t)|^2 = |\langle t(r_t) | H | \nu'
  \rangle |^2 = \sum_\alpha |V_{t\alpha}(r_t)c_\alpha^{\nu'}|^2.
  \label{eq:hopmap}
\end{equation}
In the last form the state $\nu'$ is represented in atomic orbital
basis $\alpha$.

Thus, topographic STM images can be approximated with the isosurfaces
of the square of the hopping integral. However, in the following we
shall present results for the bare hopping integral as function of
tip position --- a three-dimensional hopping map. The hopping maps
reveal the change in the phase of the current, while the constant
absolute value surfaces can be still compared to topographic STM
images. 

To calculate the hopping maps in practise, we only need the hopping
integrals between the tip and atomic orbitals of the sample, and the
weight coefficients $c_\alpha^{\nu'}$ from solving the secular equation
for the sample. Since this is extremely fast, the above method suits
for simulating very large systems. However, one has to keep in mind
that the validity of the above relationship relies on imaging the
frontier orbital with a bias voltage including only the corresponding
molecular orbital, and that the STM tip is close enough to the
molecule that the current flows mostly through the molecular state.

It is possible to get quite similar dependence for the
Tersoff-Hamann approximation for simulating STM images. In the TH
scheme surfaces of constant local density of sample states (LDOS) at
the position of the tip apex can be identified with the STM images. In
this case the LDOS is dominated by $\nu'$, so the current will be
proportional to $\langle r_t | \nu' \rangle$. While this bears close
resemblance to Eq. \ref{eq:hopmap} there may be crucial differences
especially close to the surface, as discussed in Sec. \ref{sec:resultsB}.

\section{Model And Methods}
\label{sec:model}

In this work, simulations have been carried out using our
purpose-built method for simulating STM in a TB atomic orbital basis,
and compared to results obtained using standard DFT methods (VASP
code\cite{VASP}).  In this section we first describe the structures of
the studied systems and how they have been obtained, then explain the
methods used to calculate the electronic structure of the system --
including the molecular orbitals -- and finally detail the methods
used to calculate the STM images.

\begin{figure}[h]
  \includegraphics[height = 1.5 in]{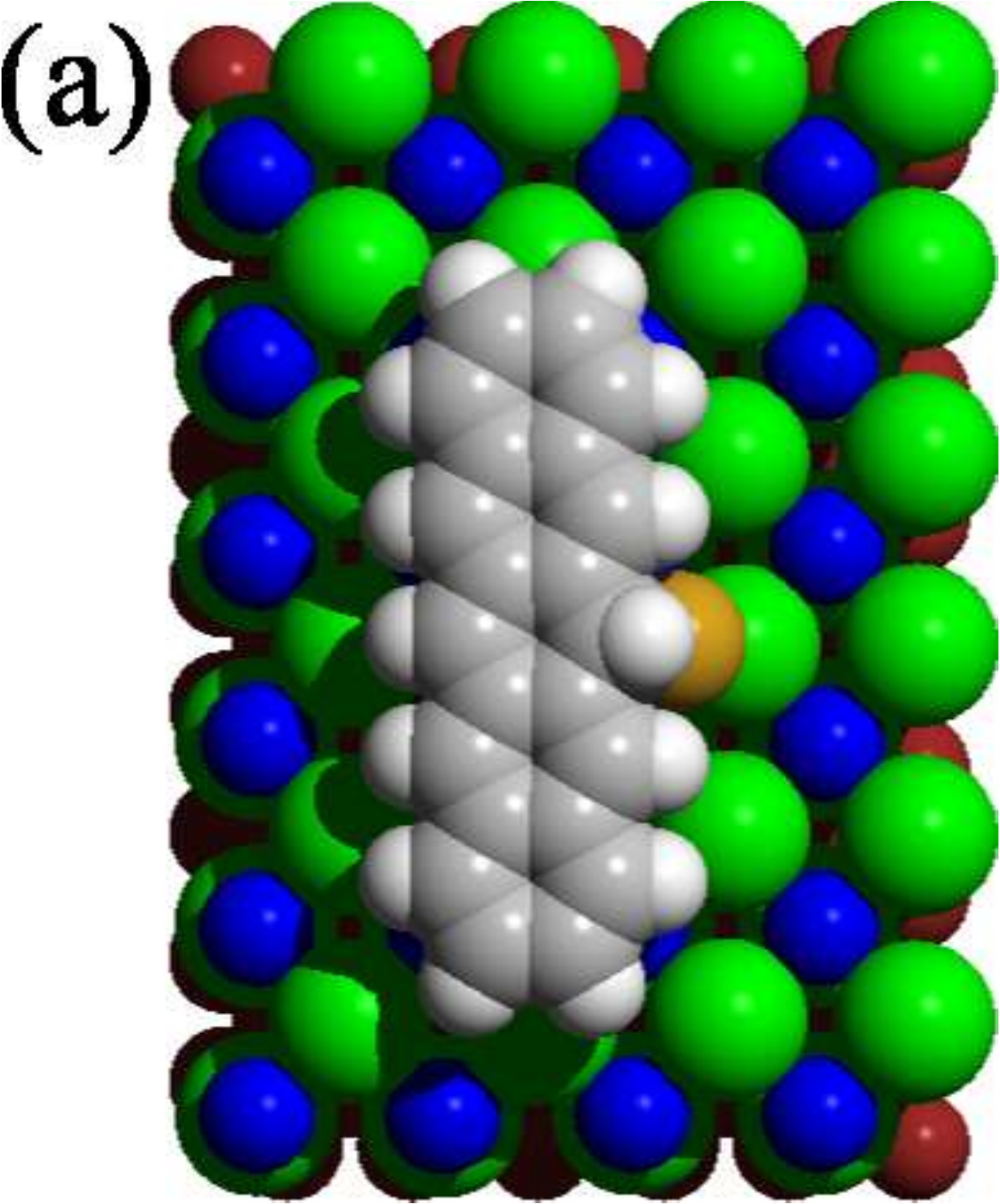}
  \includegraphics[angle=90,angle=90,angle=90,width = 1.5 in]{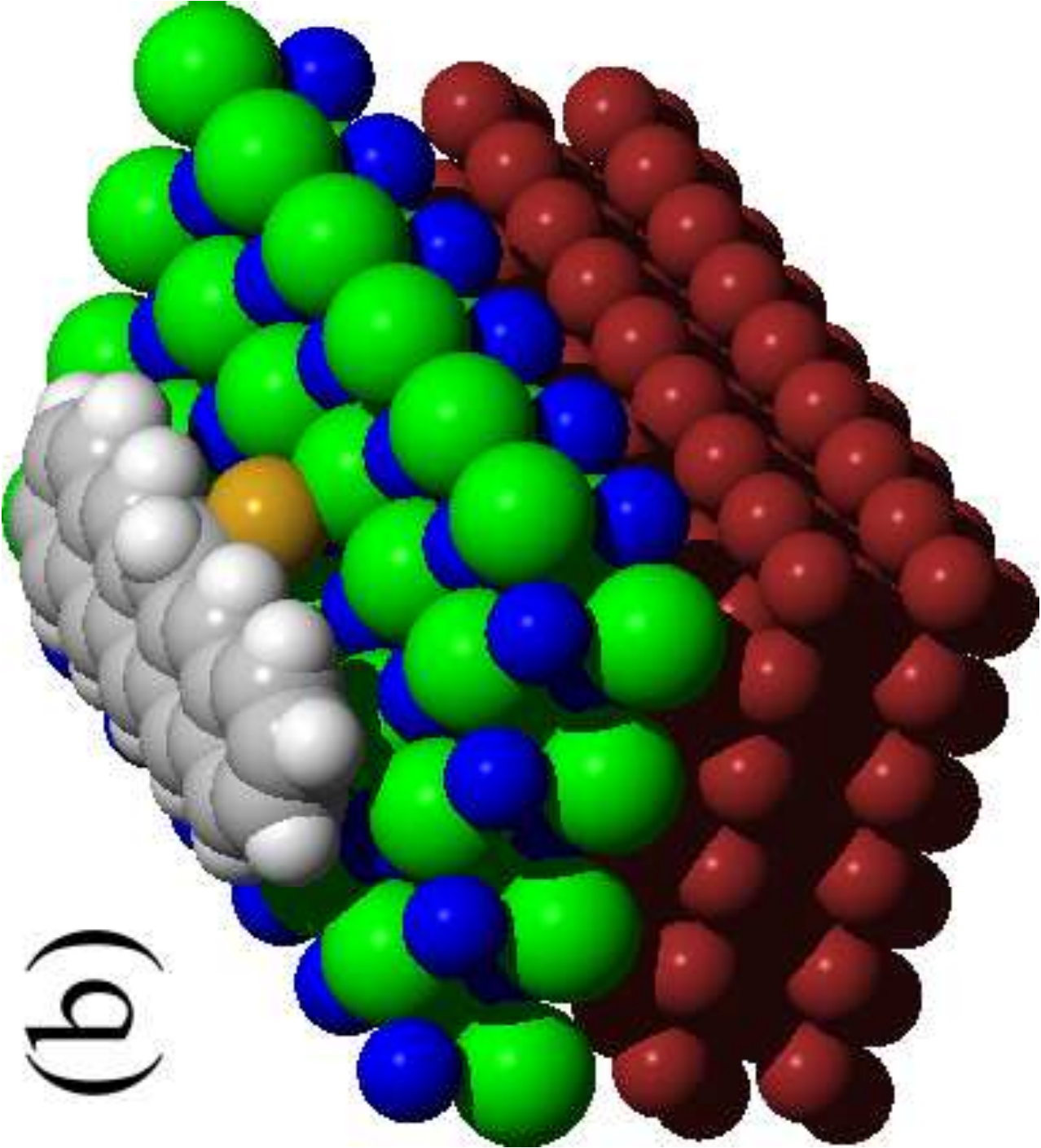}
  \caption{(Color Online) The geometric structure used in the TB
    simulations of 6-gold-pentacene on a NaCl
    -surface from two directions. Na, Cl, Cu, C, H 
    and Au atoms are shown as blue,
    green, brown, grey, white and yellow spheres, respectively.
    \label{geom_figure}}
\end{figure}

The studied system consists of a Cu(100) substrate with a bilayer of
NaCl upon which a 6-gold-pentacene complex is adsorbed. The used
geometry for the system has been obtained through DFT
calculations\cite{paavi1}. The super cell includes a slab of a four
layers of copper each layer having 6x9 atoms, and two layers of NaCl
each layer consisting of 4x6 Na and Cl ions.  In the studied geometry,
seen in Fig. \ref{geom_figure}, the gold atom is connected to the
central ring of the pentacene molecule, with the hydrogen tilting
upwards out of its way. The plane of pentacene lies 3.5 \AA\ above the
insulating overlayer, and 9 \AA\ from the topmost copper atoms. The
same geometry has been used for calculations of isolated molecules as
the adsorbed geometry differs only slightly from the relaxed isolated
molecule. In the TB calculations the STM tip has been simulated using
a pyramidal structure consisting of five copper atoms, connected to
a slab of two layers of Cu(100) surface.

Electronic structure calculations for the system, utilizing TB
formalism, have been done using angle- and distance-dependent
Slater-Koster type\cite{slat} hopping integrals $v_{\alpha \beta m}$.
The hopping integrals have been modified by including an exponential
damping to improve representation of interactions in the insulating
overlayer and the molecule.
\begin{equation}
  v'_{\alpha \beta m}(r_{ab}) = v_{\alpha \beta m} \left[
    \frac{1}{1+e^{\lambda (r_{ab}-r_0)}} + \left(1 - \frac{1}{1+e^{\lambda
          (r_{ab}-r_0)}} \right) \frac{r_{ab}^2}{r_0^2}
    e^{-\kappa(r_{ab}-r_0)}\right] 
\end{equation}
Here $\alpha$ and $\beta$ refer to atomic orbitals on atoms $a$ and
$b$, $m$ indicates the type of bond ($\sigma$, $\pi$), $r_0$
designates the distance at which the hopping integrals are changed to behave
exponentially, $\lambda$ the speed at which the change happens and
$\kappa$ the strength of the exponential damping. In this study the
values for the parameters have been: $r_0$ = 3 \AA, $\kappa$ = 1.0
\AA$^{-1}$ and $\lambda$ = 2.8 \AA$^{-1}$. A cut-off radius for the
hopping integral was used to fasten up the simulations.

The bias voltage $V$ used in the STM experiments defines the energy
range $[E_\textrm{F}, E_\textrm{F}+eV]$ and the states that can
contribute to the STM image, and thus restricts the basis set needed
in the modelling.  In all TB calculations, for C and Cl valence
electrons of s- and p-types have been included while for Cu, Au, H and
Na only valence electrons of s-type are accounted for. Au d-electrons
have been neglected since the DFT simulations indicate that the
d-orbitals of Au do not have a significant contribution near the Fermi
energy of the system and do not therefore affect the STM imaging with
reasonably small bias.  Similarly for Cu and Na only the s-orbitals
have contributions near the Fermi energy.  The tip apex Cu-atom has
also been treated as having an s-type orbital only.

In the Green's function approach, the electronic structure for the Cu
atoms in the two lowest layers of the adsorbate and for all the atoms
in tip is modelled using Haydock's recursion scheme
\cite{niemi1,Horsfield} which enables description of a larger
background and also speeds up the simulations. The two topmost layers
of Cu atoms in the adsorbate are modelled similarly to the rest of
system.

The TB calculations are carried out with a very small number of
parameters --- in addition to fixed parameters describing the hopping
integrals, the on-site energies are required for obtaining the
Hamiltonian. In general, the on-site energies are not directly fitted
to any external data but first taken from literature\cite{Harrison}
and then shifted self-consistently to keep the partial charge of each
ion neutral. This procedure is used in calculation of the free
molecules.

However, this method does not work for systems with charge transfers
between ions, and the ionic charges have to be obtained from DFT or
experiments. The on-site energies for charged Na and Cl ions were thus
adjusted to reproduce the energy-dependent density of states projected
to atomic orbitals obtained from DFT calculations. The fitting was
done only for clean NaCl bilayer on copper. The partial charges for
each Na and Cl ion were then obtained by integrating the partial
density of states.

In calculation of the combined adsorbate-substrate system the on-site
energies are again shifted self-consistently to maintain the
predefined partial charges (zero for all other ions except the ions in
the polar film, for which the values obtained for the clean film is
used). Thus there are actually no parameters to be fitted for the
combined system, which increases the prediction power of the method.

Previously, the functional forms of the hopping integrals used in the
TB method had been tested for the clean NaCl bilayer and free organic
molecules. The obtained molecular orbitals accurately match DFT
Kohn-Sham states in shape and qualitatively in energy for
chloronitrobenzene\cite{niemi2} and
pentacene\cite{akorunpub}. Finally, because DFT calculations with and
without spin-polarization gave similar results for STM simulations,
the TB calculations were done without taking spin into
account\cite{spinendnote}.

The calculation of the tunneling current using the TB method is done
according to the Todorov-Pendry approach. For comparison, the STM
calculations using TB have been done using two kinds of functional
forms for the hopping integral between the tip and the sample. The
first is the same modified Slater-Koster type\cite{slat} hopping which
is used in the electronic structure calculations, with exponential
damping. The second method is through the Wolfsberg-Helmholz
approximation using overlap integrals for Slater-type atomic orbitals
(STO), similarly to extended Hückel theory
(EHT).\cite{hoffmann,anderson} These hopping integrals have the form:
\begin{equation}
  v'_{\alpha \beta}(r_{ab}) = \frac{1}{2}k S_{\alpha \beta} \left(
  H_{\alpha \alpha} + H_{\beta \beta} \right),  
\end{equation}
where $k=1.75$. The overlaps $S_{\alpha \beta}$ are
calculated using the well-known analytical forms presented by Mulliken
\textit{et al.}\cite{mulli49}. The Clementi-Raimondi\cite{clera} screening
constants were used for all species of atoms. The $H_{\alpha \alpha}$
and $H_{\beta \beta}$ elements are the ionization energies
for the atomic orbitals in question. 

\section{Results}
\label{sec:results}

\subsection{Molecular Orbital Decompositions} 
\label{sec:resultsA}

Molecular orbital calculations can provide a powerful tool in the
analysis of STM results on insulating overlayers since the
conductivity of molecules adsorbed on such surfaces is usually heavily
influenced by the frontier molecular orbitals\cite{repp}. The
gold-pentacene complex has a singly occupied molecular orbital (SOMO)
which dominates the STM current on small biases. In
Ref. \onlinecite{paavi1} the SOMO state was argued to have significant
contributions from the highest occupied molecular orbital (HOMO) and
lowest unoccupied molecular orbital (LUMO) of pentacene, in addition
to the gold 6s-orbital.

In Table \ref{somo_weights} can be seen the weight coefficients
$c_{\mu}^{i}$ of free pentacene molecular orbitals, and the gold
6s-orbital to the SOMO state of the Au-pentacene complex.
Interestingly it can be seen that the SOMO state is formed as a linear
combination of both the HOMO and LUMO orbitals with almost equal
weights, with the gold s-orbital also contributing greatly. This also
verifies the assumptions of Ref. \onlinecite{paavi1} made using
geometrical arguments. From the square sum of these coefficients
(0.91) can be deduced that the other pentacene states that lie near
the Fermi energy, have only a very small contribution to the SOMO. The
states above and below in energy to the SOMO (designated here as
SOMO-1 and SOMO+1) can be seen to originate from the pentacene LUMO
and HOMO states, respectively, with a large contribution from the gold
6s orbital and a smaller one from other pentacene states.

\begin{table}[h]
  \centering
  \begin{tabular}{|c|c|c|c|}
      \hline
      $c_{\mu}^{i}$& SOMO$-$1 & SOMO & SOMO$+$1\\
      \hline
      HOMO & 0.69 & 0.61 & 0.20\\
      \hline
      LUMO & 0.25 & 0.66 & 0.59\\
      \hline
      Au$_\textrm{6s}$ & 0.34 & 0.31 & 0.62\\
      \hline
    \end{tabular}
  \caption{The contributions of the HOMO and LUMO of the free
    pentacene molecule and the 6s orbital of the gold atom to the SOMO
    state which is formed as Au bonds with pentacene. The values have 
    been calculated according to Eq. \ref{coeffeq} in Sec. \ref{sec:theory}.
    \label{somo_weights}} 
\end{table}

\begin{figure}[h]
  \includegraphics[height = 1.4 in]{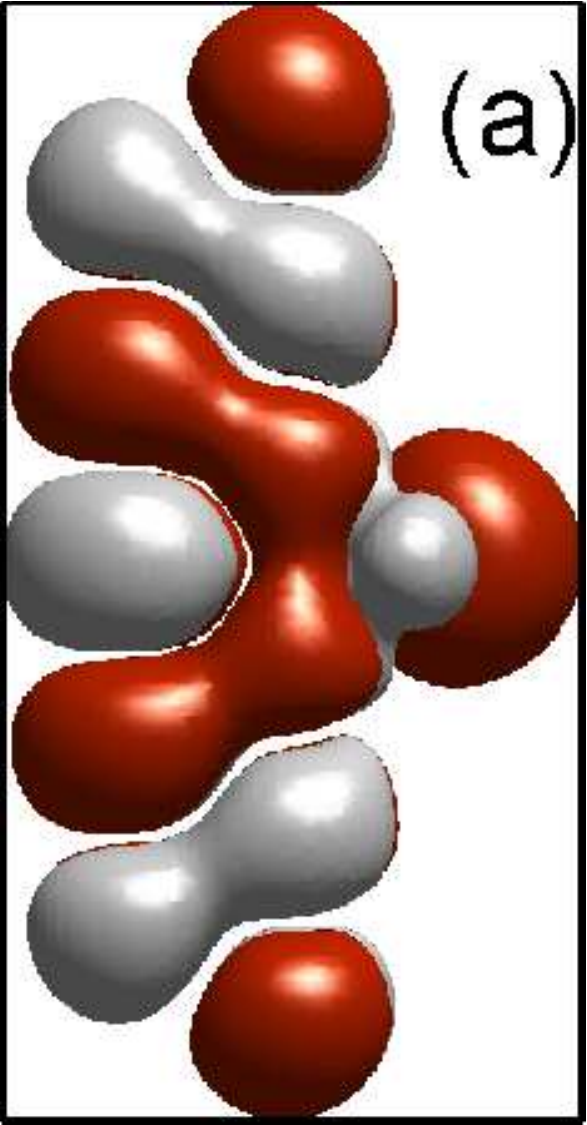}
  \includegraphics[height = 1.4 in]{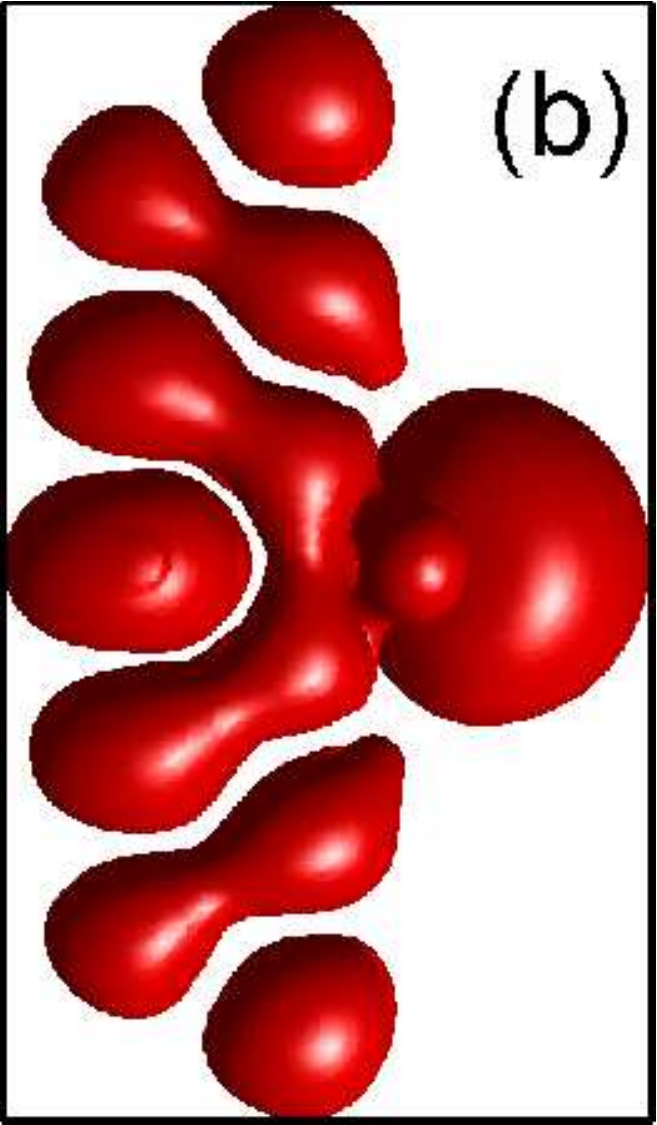}\\
  \includegraphics[height = 1.4 in]{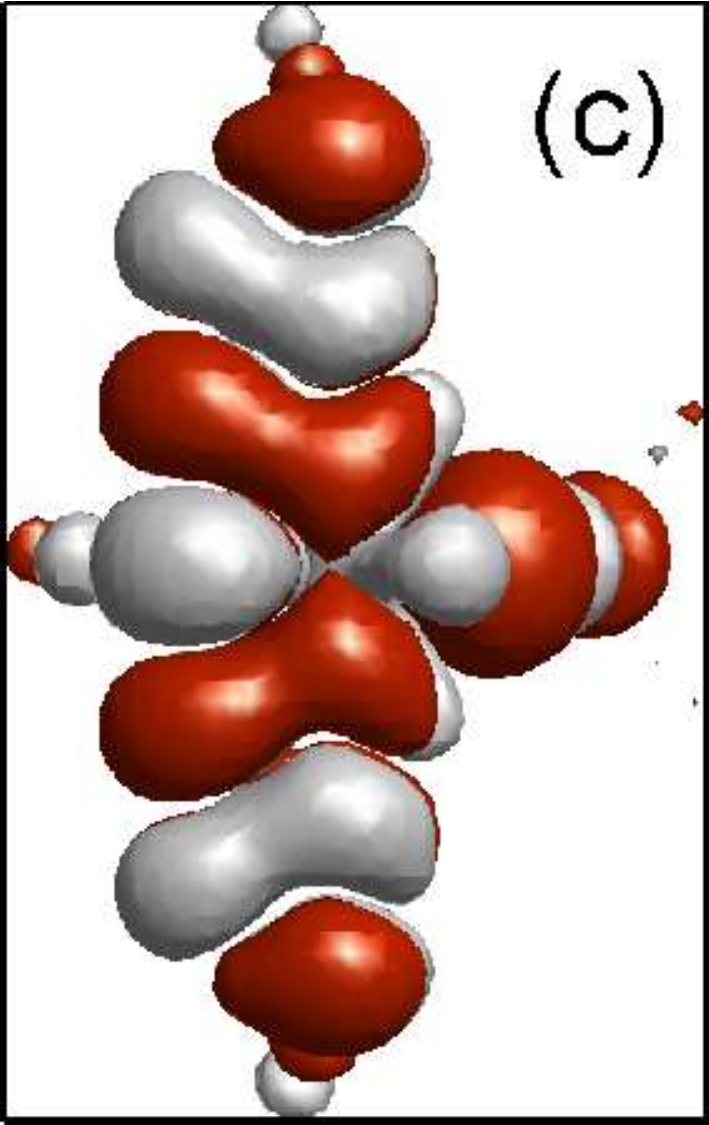}
  \includegraphics[height = 1.4 in]{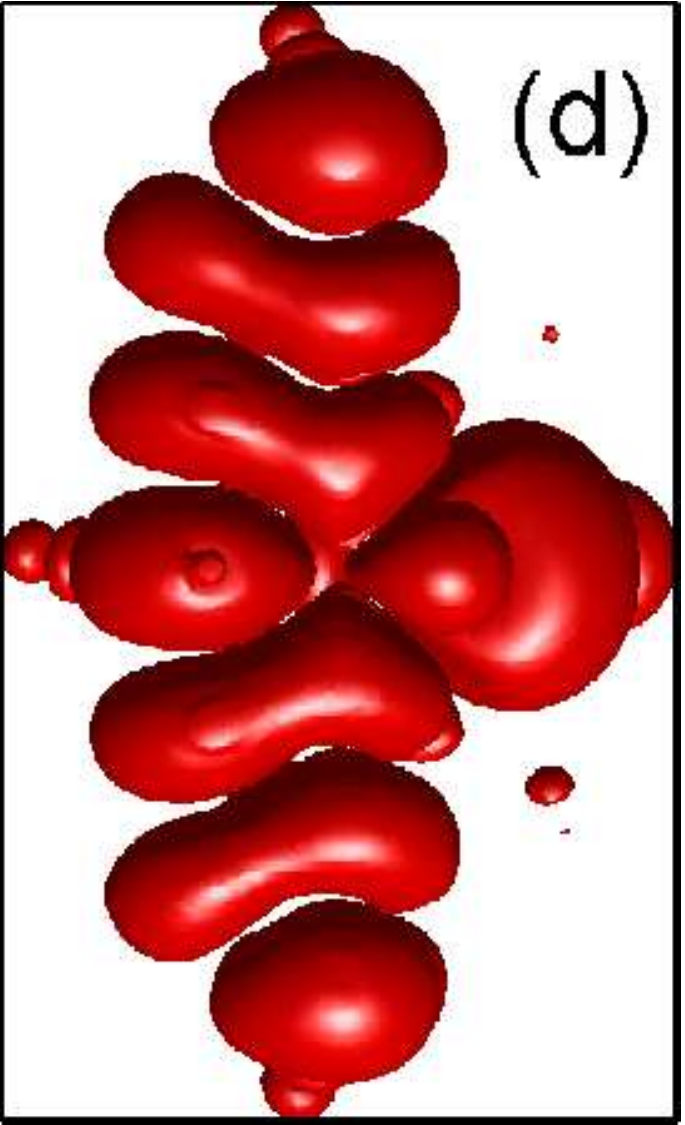}
  \caption{(Color Online) Isosurfaces of molecular orbitals of free Au-pentacene
    complex (SOMO) calculated using (a) TB and (b) DFT, and adsorbed
    Au-pentacene complex (SOMO$^{*}$) with (c) TB and (d) DFT.   
    In the TB molecular orbitals
    the lobes with different signs have been plotted using red and
    white. The molecular orbitals in the DFT calculations correspond to single
    Kohn-Sham states. 
    \label{molorbs}}
\end{figure}

The TB and DFT calculations for SOMO molecular orbital of a free gold-pentacene
system (as seen in Fig. \ref{molorbs} (a) and (b)) are found to be in
good accordance with each other, testifying of a
successful fitting for the electronic structure in the TB method. The only
notable difference between the results is that the gold s-orbital is
distinctly contracted in the TB calculations compared to the DFT
results. In the free molecule calculations there is a bridge combining
two orbital lobes with equal signs, clearly seen in the center of
Figs. \ref{molorbs}(a) and (b). The bridge runs along the
long axis of the molecule, seen vertically in the pictures.

In Fig. \ref{molorbs}(c) and (d) is presented the molecular orbital
calculations for the corresponding SOMO$^{*}$ state of the system
consisting of the adsorbed molecule, the insulating overlayer and the
copper substrate. Similarly to the free molecule calculations, the
results are in good accordance with each other except that the gold
s-orbital still being distinctly contracted in the TB calculations
compared to the DFT.  In the adsorption, the bridge seen in the free
molecule calculations has disappeared entirely, leaving a cross-shaped
nodal area. The bridge is also absent in experimental STM images of
the SOMO$^{*}$ state.

This absence of the bridge happens due to interaction of the molecule
with the chlorine atom which lies on the axis connecting gold with its
nearest carbon. Simulations done using both VASP and TB (with the
chlorine interacting solely with Au) using a system consisting of the
chlorine in question and the Au-pentacene complex exhibit the same
behaviour of the bond breaking, as the atom is brought closer to the
molecule from further away. A similar shift in energy is found for the
chlorine p-orbital using either DFT or TB with predefined charge
maintenance. In effect, this can be understood as a sign that some
interactions between adsorbed molecular systems and the insulating
overlayer can affect the molecular states, and therefore also the STM
images which are integrally connected to them.

\subsection{STM simulations}
\label{sec:resultsB}

\begin{figure}[h]
  \includegraphics[height = 1.4 in]{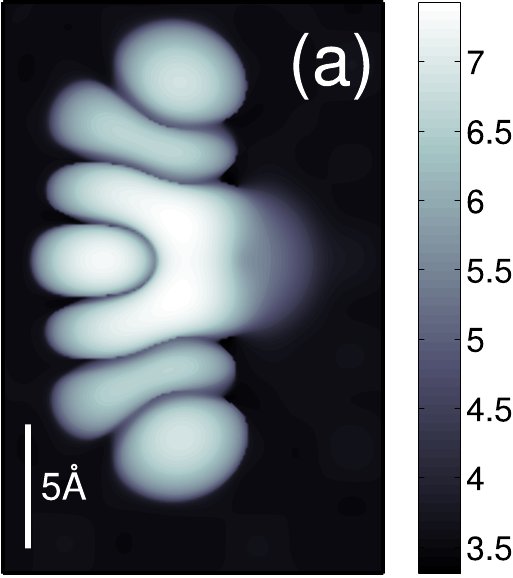}
  \includegraphics[height = 1.4 in]{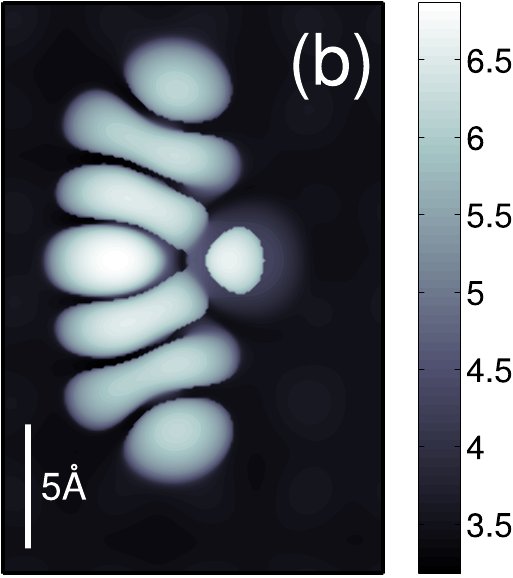}\\
  \includegraphics[height = 1.4 in]{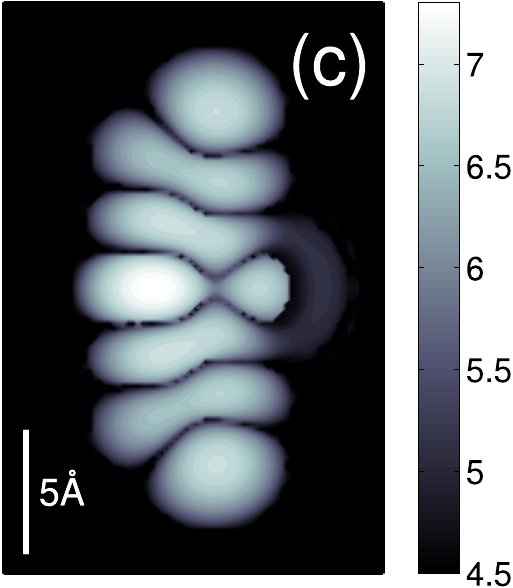}
  \includegraphics[height = 1.4 in]{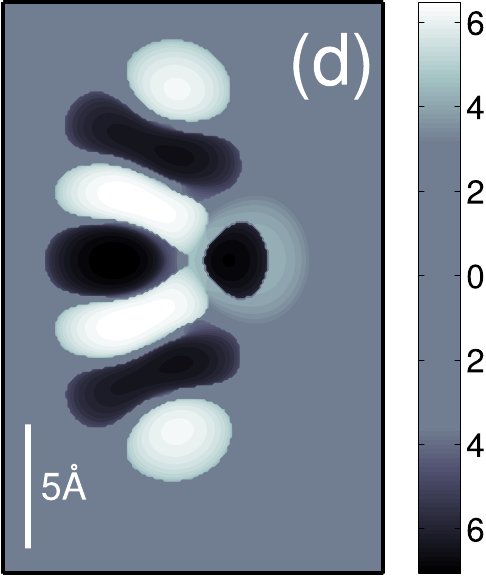}

  \caption{(Color Online) Simulated STM images of the 6-gold-pentacene
    complex calculated using (a) TB-TP with Slater-Koster type hopping
    integrals, (b) TB-TP with STO-EHT type hopping integrals and (c)
    DFT-TH. In (d) is shown a hopping map of the adsorbed
    molecular orbital calculated using STO-EHT hopping integrals. In all
    images the height corresponds to the distance from the topmost insulating
    overlayer. In (d) the color axis is extended in two
    directions to visualize the phase information.  The background in the
    hopping map has been set to zero to clarify the image.
    \label{top_stm}}
\end{figure}

The constant current STM images showing the SOMO$^{*}$ orbital
simulated using TP approach of Eq.\ref{eq:TPcurrent} are shown in
Figs. \ref{top_stm}(a) and (b) calculated with the two different
hopping integrals between the STM tip and the adsorbate. For
comparison, in Fig.  \ref{top_stm}(c) is shown the corresponding
simulations carried out using DFT within TH approximation. The STM
images calculated with DFT and STO-EHT hopping integrals can be seen to be
similar to each other, while the image calculated with the
Slater-Koster hopping integrals clearly differs from them --- especially near
the bridge area. The DFT and STO-EHT images also clearly resemble the
experimental STM results\cite{paavi1} and the molecular orbital
calculated for the adsorbed molecule (see Fig. \ref{molorbs}).

Figure \ref{top_stm} (d) presents a plotting of the hopping map to the
SOMO$^{*}$ orbital using STO-EHT hopping integrals. The presented image
corresponds to a constant current STM image.  To clarify the revealed 
phase information, the sign of the
hopping integral is visualized by extending the color axis in two
directions.  The background, which would go to minus infinity, is set
to zero to help interpreting the image. The presented hopping map is
nearly identical to the corresponding STM image, due to the SOMO$^{*}$
state dominating the STM image, and it is of the order of 100 times faster to 
calculate than the TP image of Fig. \ref{top_stm}(b).

Both TP with STO-EHT and the hopping maps reproduce nicely the arc of
the experimental STM images around the Au atom.  Similarly to the
molecular orbitals, presented in Fig. \ref{molorbs}, the arc
originating from the Au 6s-orbital is slightly smaller in the TB
calculations compared to DFT. This is due to either the too weak
interaction strength between the Au s-orbital and the molecule or too
small extension of the the Au Slater orbital used are calculated in
the electronic structure calculations. Notice that no especial fitting
was done for the calculation but the parameters for Au are directly
taken from literature.  The arc is visible in the TB images only close
to the surface and thus the TB and DFT calculations are done at
slightly different distances.

STM images for TB have been calculated with bias 0.08 $V$ and for DFT
with 0.5 $V$\cite{spinendnote}. One should note, that the theoretical
values are not directly comparable with experimental voltages, as the
Coulomb blockade in the experiments shifts the observed energies of
the states\cite{paavi2}. The bias dependence of the STM image is small
for both modelling methods with reasonably low biases. The
corrugations of the images can be seen to be very similar in all
images.  The TB images differ in size from the DFT images, due to the
lower imaging distance.

\begin{figure}[h]
  \centering
    \includegraphics[width = 2.6 in]{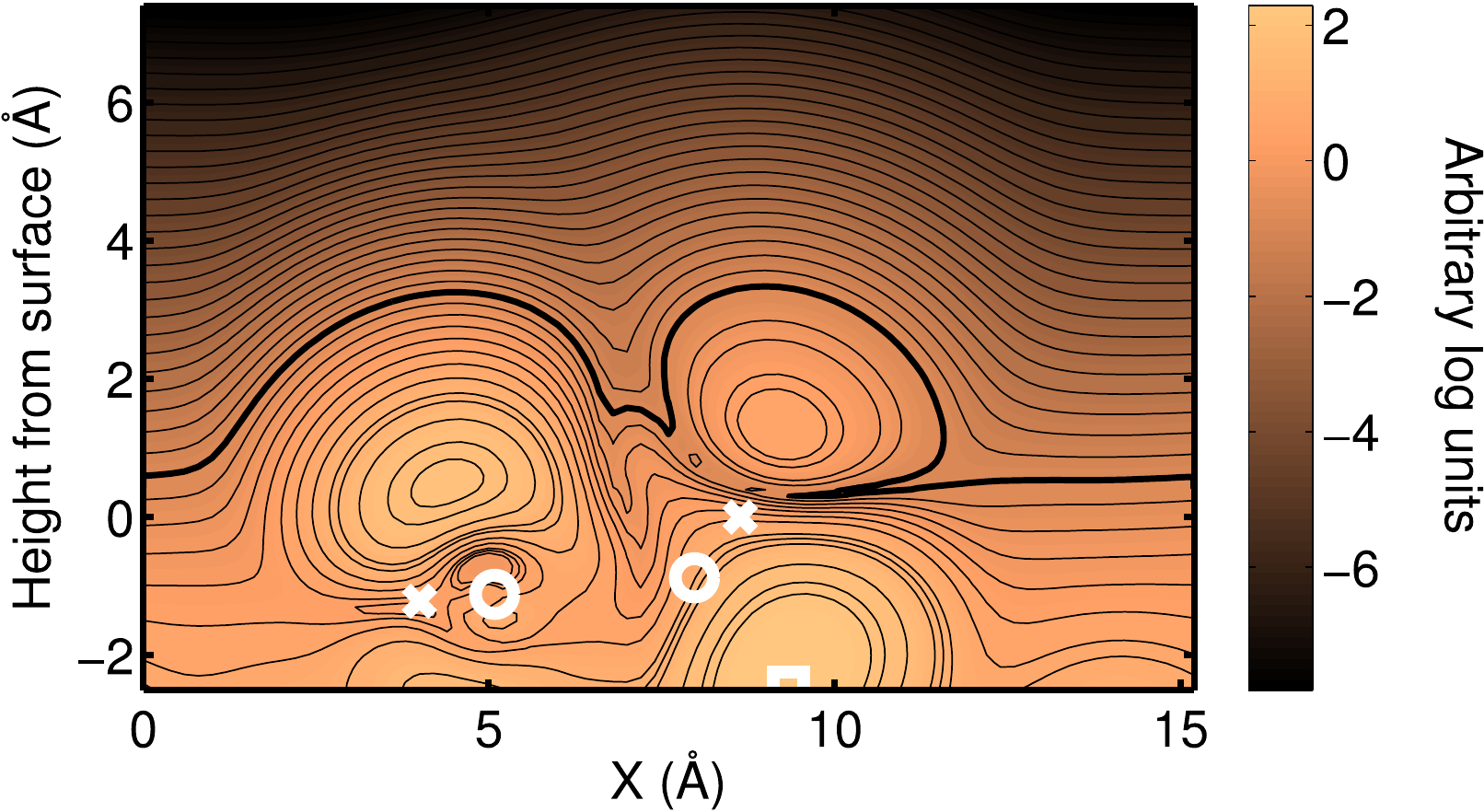}
  \caption{(Color Online) Logarithmic plotting of the STM current as a
    function of height in the plane of the Au-Pentacene bond,
    calculated using TB with STO-EHT hopping integrals. Positions of carbons in
    the plane marked with crosses, hydrogens with circles and gold
    with a square. Lines correspond to constant current surfaces. An
    ambiguous surface is highlighted.  \label{logside_stm}}
\end{figure}

Even though there is little bias dependence for the STM image of the
system, the image may be quite dependent on the chosen
constant current surface, i.e. the initial height from the adsorbate. To
visualize this we have calculated the STM current in the plane where
the C-Au bond lies. The result is shown in Fig. \ref{logside_stm} as a
logarithmic mapping of the STM current as a function of height. 
There are several interesting properties to the STM image, which can
be seen from this figure. The most striking property is the fact that
the STM current is clearly not unambiguous; there exist folded
constant current surfaces --- a predefined current does not give a
unique height. This might have
implications in experiments, such as that the scanning direction
affects the outcome of the measurement.

One should note, that according to the geometry obtained in the DFT
calculations, in this system the highest atom lies 4.5 \AA\ above the
insulating overlayer, and approximately 10 \AA\ from the atoms in the
conducting surface. Because of this, the distance between the STM tip
and the molecule cannot be very large while still being able to
generate a measurable tunneling current. As can be seen in
Fig. \ref{logside_stm}, an ambiguity can be found also quite far away
from the molecule itself. Thus, it is possible that the experiments
are done within a range where the above multiplicity can occur.
However, being close to the molecule implies that the actual effect on
experiments will depend on the real structure of the STM tip.

In addition to this, the contrast and extension of the STM image
clearly depend on the followed constant current surface. As is
well-known, the STM image is seen to widen further away from the
molecule, and the contrast of the nodal region in the middle vanishes
at higher distances. However, due to complicated shape of the SOMO
orbital, the topographical STM image may now be also qualitatively
different with various values of the tunneling current. Thus, this may
give an answer to the experimentally observed phenomena of the arc
appearing in some experimental STM images while being absent in
others. From the logarithmic mapping, as well as the calculated STM
images with different initial heights, can be interpreted that the arc
should be seen at low scanning heights but disappear when going
further away from the surface. 

\begin{figure}[h]
  \centering
  \includegraphics[width = 2.6 in]{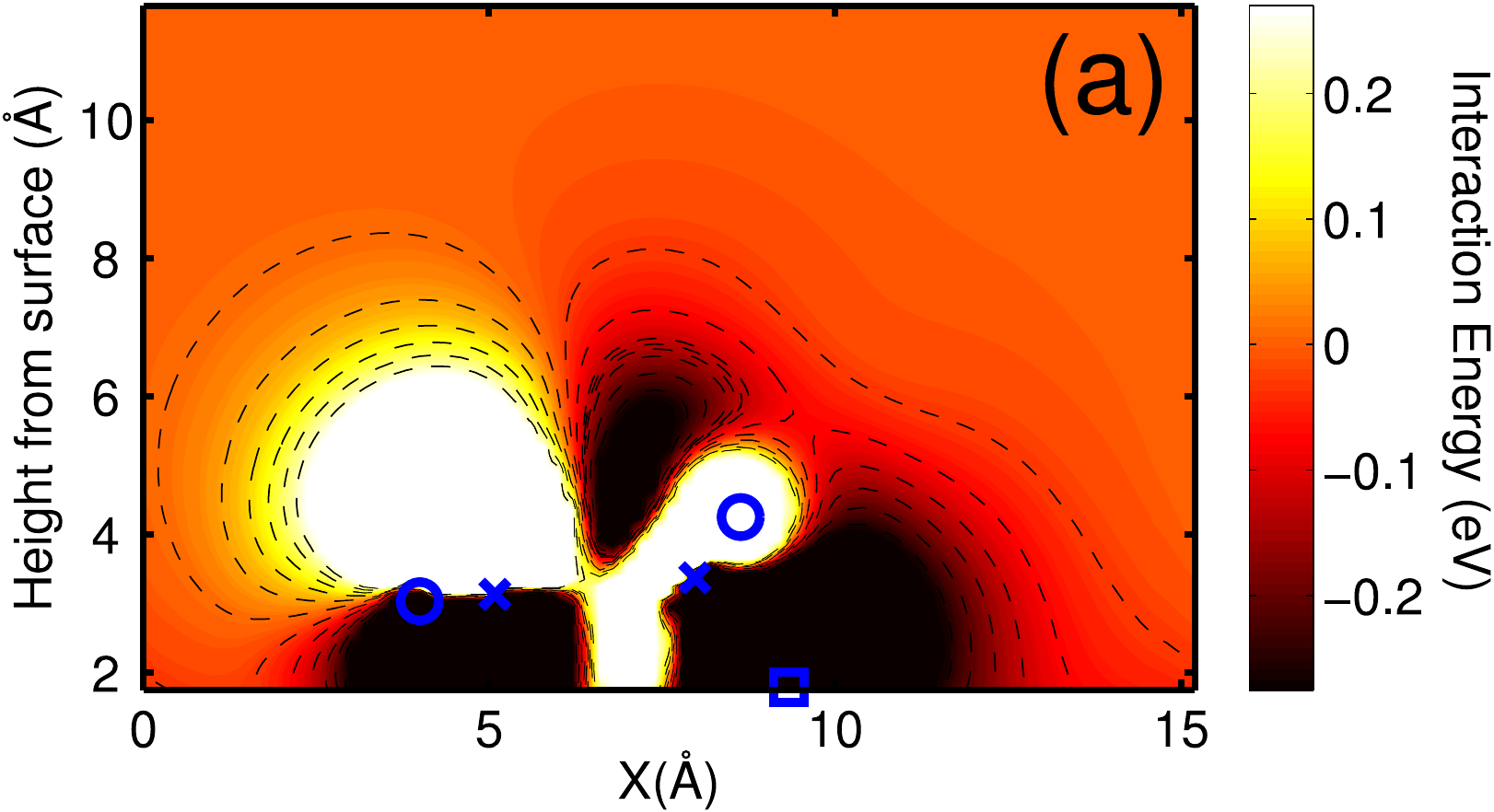}\\
  \includegraphics[width = 2.6 in]{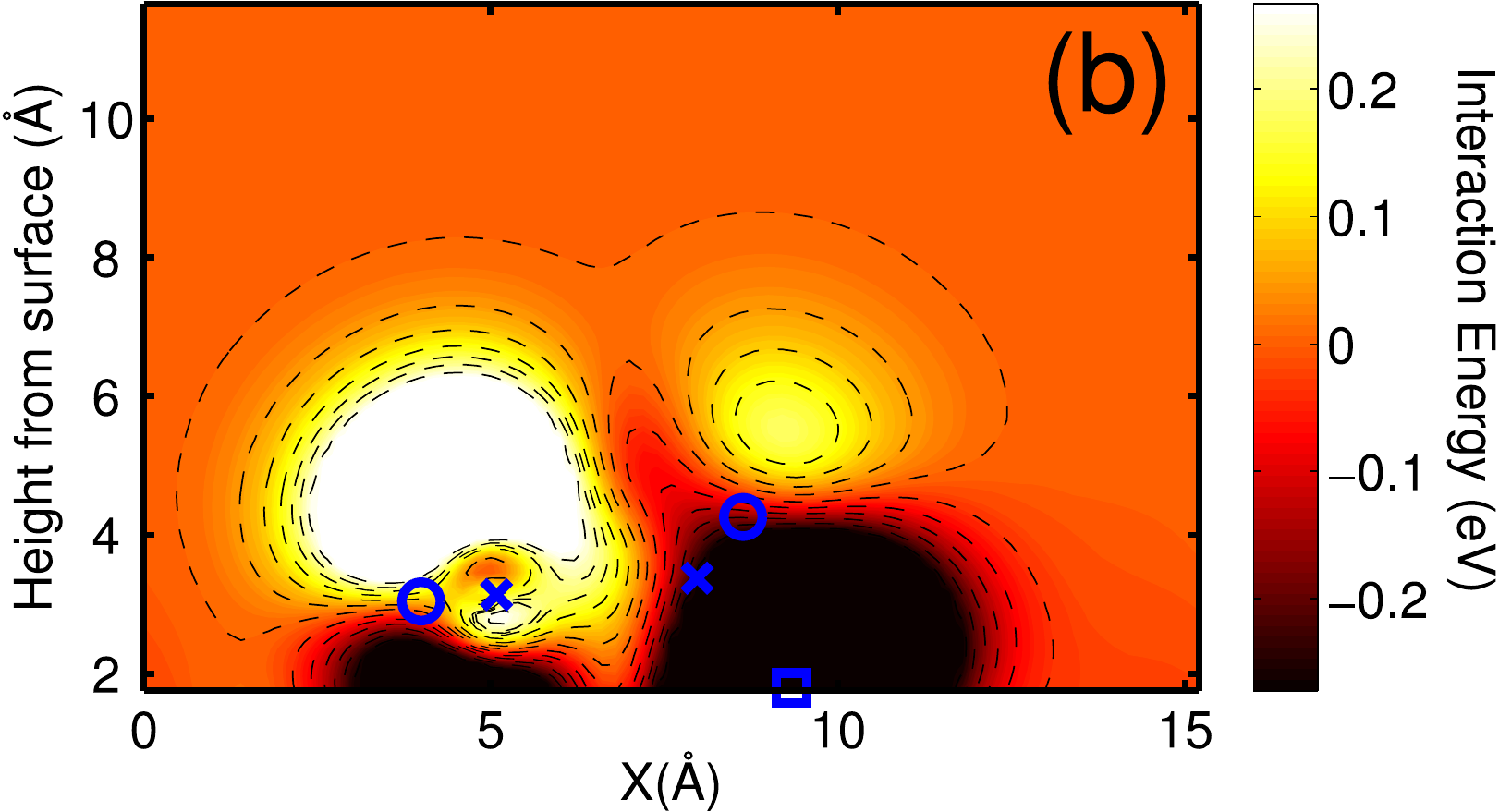}\\
  \includegraphics[width = 2.6 in]{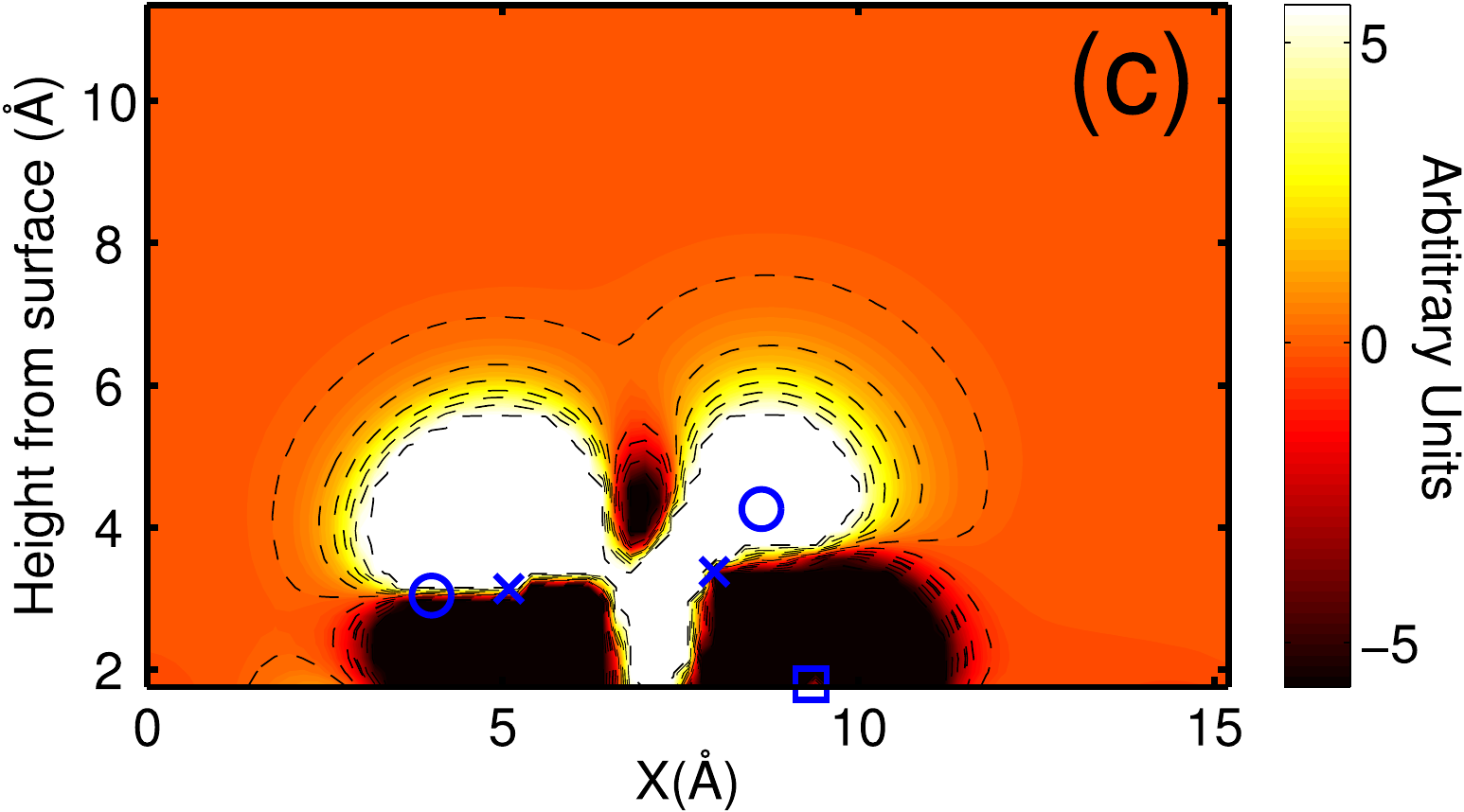}

  \caption{(Color Online) Images of the spatial dependence of the
    (a) Slater-Koster type hopping integrals to the SOMO$^{*}$ orbital, (b)
    STO-EHT type hopping integrals  to the SOMO$^{*}$ orbital and (c) the
    local density of the SOMO$^{*}$ orbital (calculated using TB) as a function of
    height in the plane of the Au-pentacene bond. Locations of carbons
    in the plane marked with crosses, hydrogens with circles and gold with a
    square. The values of the hopping integrals and the
    molecular orbital have been truncated to help interpretation of the
    image.  \label{hopside}}
\end{figure}

The STM images calculated using the two hopping integrals
(Fig. \ref{top_stm} (a) and (b)) emphasize the importance of the
proper choice of the hopping integral in TB calculations. The differences
between the two approximations can be further visualized with the
hopping maps whose cross sections along the plane of the Au-C bond are
shown in Fig. \ref{hopside} (a) and (b) for the used Slater-Koster and
STO-EHT type hopping integrals, respectively. The hopping map results
differ drastically. Where in the STO-EHT a slight negative protrusion
rises slightly between the two positive lobes, in the second image it
is a large negative lobe that encases the positive lobe by connecting
to the negative lobe centered around the gold atom. This is the reason
for the clear differences between the STM images for the two methods
in Fig.  \ref{top_stm}. This is partly due to the incorrect spatial
dependence of the Slater-Koster hopping integrals neglecting the size
(principal quantum number) of the orbitals. This is especially
important in this kind of a system in which a heavy atom (gold and its
6s-orbital) plays a major role. On a brighter note, for this kind of
systems the STO-EHT hopping integral gives a good accordance with both
DFT and experimental results.

For comparison, a similar cross-section of the SOMO$^{*}$ state is
shown in Fig. \ref{hopside}(c). Even though similarity with STO-EHT
hopping map further away from the surface, there are some
discrepancies close to the molecule. This highlights the difference in
Tersoff-Hamann approach compared to Todorov-Pendry approximation.

By modifying the range at which the exponential tail is applied and by
changing the relative differences of s-p and s-s interactions the
Slater-Koster type hopping integrals can be brought into line with the
DFT and STO-EHT results, producing similar STM images and hopping
maps. However, this greatly affects the ability of the method to
predict measured STM images, especially compared to the STO-EHT, for
which no specific fitting was carried out. This emphasizes the
importance of a proper modelling of the tip-adsorbate interaction.

\section{Summary}
\label{sec:summary}

The objective of this work has been to develop TB based methods for
simulating STM images of complicated molecules on insulating
overlayers. We used pentacene bonded with gold on a sodium chloride
thin film with copper substrate as a test system. Results obtained using our
purpose-built method for simulating STM images are found to be
sensitive to the choice of hopping integral between the STM tip and
the adsorbate.  

With STO-EHT -type hopping integrals the STM results very closely
match both experiments and DFT simulations. The only notable
difference found is regarding the size and visibility of the arc
formed by the Au 6s-orbital. Instead, the Slater-Koster -type hopping
integrals give clearly differing results to both other simulations and
experiments, due to their disregard for the size of the
orbitals. 

A logarithmic plotting of the STM current with the STO-EHT hopping
integrals predicts that the complex structure of the molecular
orbitals may lead to ambiguity of the STM image. This kind of effect
may have implications in experiments, such as the STM image changing
with respect to the scanning direction. Additionally, the tip may
drift too close to the molecule in constant current imaging, allowing
the molecule to jump to the tip. The actual effects of the ambiguity
will largely depend on the actual shape of the tip, which would merit
a further study.  These results highlight the importance of combining
experiments with simulations in order to properly interpret the
obtained results.


Hopping maps, for which the theory has been presented in this work,
provide a powerful tool, not only in their ability to study the phase
of the STM current, but also in the fact that they can be used to very
easily simulate the STM results of different conformations even for
large systems. As can be seen from the results presented here, for
systems where a single molecular orbital dominates the STM image, the
hopping map is very similar to the STM image. The phase of the STM
current can be utilized to analyze the formation of the STM current
from different parts.

The methods developed in this work will be used 
to study large molecules suitable for molecular
electronics on insulating overlayers. The ability to simulate large
systems with comparably small computational resources, especially
when utilizing the hopping maps, makes it more feasible to simulate
relevant single-molecular systems. 

\appendix
\section{Different formulations of Todorov-Pendry approach}
\label{sec:appendix}

The current running through a system is obtained in
Todorov--Pendry\cite{todorov} approach from
\begin{equation}
\label{eq_currentB}
I = \frac{2 \pi e}{\hbar} \int_{E_F}^{E_F+eV} \,\mathrm{Tr} [\rho^{0}_{\tau
    \tau'}(E-eV) T_{\tau' \sigma'}^{\dag} \rho^{0}_{\sigma'\sigma}(E)
  T_{\sigma \tau}]dE.
\end{equation}
In tight-binding basis, the density matrix $\rho$ is easily calculated with
the aid of Green's function,
\begin{equation}
\label{eq_dosB}
\rho_{if}(E)=\frac{\eta}{\pi} \sum_s G_{is}^+ G_{sf}^{-}  ,
\end{equation}
where $i$, $f$ denote initial (for example, in the tip) and final state
(in the sample).

With the aid of Eqs. \ref{eq_currentB} and \ref{eq_dosB} it is easy to 
transform the integrand into form
\begin{equation}
j =  \frac{2 \pi e}{\hbar} \,\mathrm{Tr} [ \rho_{\tau '\tau }^0
  T^\dagger_{\tau \sigma } \rho_{\sigma \sigma '}^0 T_{\sigma '\tau
    '}] 
=\frac{2 e \eta }{\hbar}  \,\mathrm{Tr} [\rho^{0}_{\tau' \tau } 
T^\dagger_{\tau \sigma } G^{0-}_{\sigma f} G^{0+}_{f
    \sigma' } T_{\sigma' \tau' } ].
\label{currentT}
\end{equation}

Various other formulations to calculate tunneling current can be
recovered from the equation above.  If we apply \ref{eq_dosB} once
again, we get the following form:
\begin{displaymath} 
j  =\frac{4 e \eta^2 }{h}  \,\mathrm{Tr} [ G^{0+}_{i
    \tau }  
T^\dagger_{\tau \sigma } G^{0-}_{\sigma f} G^{0+}_{f
    \sigma' } 
T_{\sigma' \tau' }G^{0-}_{\tau' i} ]  \nonumber
\end{displaymath}

Assuming that tip and sample are disconnected in the beginning and 
utilizing Dyson's equation:
\begin{displaymath}
G^{-}_{i f} = G^{0-}_{i \tau } T^\dagger_{\tau \sigma } G^{0-}_{\sigma f} =
G^{0-}_{i \tau } V^\dagger_{\tau \sigma } G^{-}_{\sigma f}, 
\end{displaymath}
since there are no tip-sample matrix elements in the case of uncoupled system.

Thus it is straightforward to convert Eq. \ref{currentT} to the form
\begin{equation}
j  =\frac{2 e \eta }{\hbar}  \,\mathrm{Tr} [\rho^{0}_{\tau' \tau } 
V^\dagger_{\tau \sigma } G^{-}_{\sigma f} G^{+}_{f
    \sigma' } V_{\sigma' \tau' } ],
\label{currentV}
\end{equation}
which is essentially the formalism used in this work. 

There are some extensions and alternative formulations that can be
shown here.  First, the convergence parameter $\eta$ can be
generalized to an energy dependent broadening parameter $\Gamma$/2
related to self-energy.  Hence we can express this in a
Landauer-B\"uttiker form
\begin{equation}
j  =\frac{e}{\hbar}  \,\mathrm{Tr} [\rho^{0}_{\tau' \tau } 
V^\dagger_{\tau \sigma } G^{-}_{\sigma f'} \Gamma_{f'f} G^{+}_{f
    \sigma' } V_{\sigma' \tau' } ]  
\label{currentLB}
\end{equation}
as formulated by Meir and Wingreen\cite{MW}.

In addition, applying Dyson's equation leads to the equation
\begin{displaymath}
I =\frac{4 e \eta^2}{h}  \,\mathrm{Tr} [ G^{-}_{i f} G^{+}_{f i}]
\end{displaymath}
which is the result given by Ness and Fisher\cite{NessFisher}.

One of the equivalent formulations is given by McKinnon and Choy
\cite{McKinnon}, which has been written in a more elegant form by 
Mingo \textit{et al.}\cite{Mingo}.
 A further application of Dyson's equation to Eq. \ref{currentT} tells us
\begin{displaymath}
\left \{
\begin{array}{ccc}
G^{-}_{if}& =&  G^{-}_{ii'}V^{\dagger}_{i'f'}G^{0-}_{f'f}\\ 
G^{-}_{i'i}& =& G^{0-}_{i'i} + G^{-}_{if''}V^{\dagger}_{f''i''}G^{0-}_{i''i} 
\end{array}
\right.
\end{displaymath}
which means 
\begin{displaymath}
G^{-}_{ii}=G^{0-}_{ii'}(I - V^{\dagger}_{i'f'}G^{0-}_{f'f'}V^{\dagger}_{f''
i''}G^{0-}_{i''i'})^{-1}
\end{displaymath}
and
\begin{displaymath}
G^{-}_{if}=G^{0-}_{ii'}(I - V^{\dagger}_{i'f'}G^{0-}_{f'f'}V^{\dagger}_{f''
i''}G^{0-}_{i''i'})^{-1}V^{\dagger}_{i'f'}G^{0-}_{f'f} = G^{0-}_{ii'}
T^{\dagger}_{i'f'}G^{0-}_{f'f}.
\end{displaymath}
Following Mingo \textit{et al.}\cite{Mingo} we utilize a shorthand notation
\begin{displaymath}
T^{\dagger}_{i'f'} = (I - V^{\dagger}_{i'f'}G^{0-}_{f'f'}V^{\dagger}_{f'' i''}G^{0-}_{i''i'})^{-1}V^{\dagger}_{i'f'}
= D^{-}_{i'i'}V^{\dagger}_{i'f'}
\end{displaymath}

Which leads to the form:

\begin{equation}
j =\frac{2 e \eta }{\hbar}  \,\mathrm{Tr} [\rho^{0}_{\tau' \tau } 
 D^{-}_{\tau \tau}V^\dagger_{\tau \sigma } G^{0-}_{\sigma f} G^{0+}_{f
    \sigma' } D^{+}_{\sigma' \sigma'}V_{\sigma' \tau' } ]  =
\frac{2 \pi e}{\hbar}\,\mathrm{Tr} [\rho^{0}_{\tau' \tau } 
 D^{-}_{\tau \tau}V^\dagger_{\tau \sigma } \rho^{0}_{\sigma 
    \sigma' } D^{+}_{\sigma' \sigma'}V_{\sigma' \tau' } ].
\label{currentD}
\end{equation}

Now it mostly depends on the computational method, which one is the most suitable 
formalism to calculate the tunneling current.

\bibliographystyle{apsrev}


\begin{thebibliography}{20}

\bibitem{repp2} J.~Repp, G.~Meyer, F.E.~Olsson, and M.~Persson, Science
  \textbf{305}, 493 (2004).

\bibitem{repp} J. Repp, G. Meyer, S. Stojkovi\'c, A. Gourdon, and
  C. Joachim, Phys. Rev. Lett. \textbf{94}, 026803 (2005).

\bibitem{paavi1} J. Repp, G. Meyer, S. Paavilainen, F. Olsson and
  M. Persson, Science \textbf{312}, 1196 (2006).

\bibitem{nieminen} J.A.~Nieminen and S.~Paavilainen, Phys.~Rev.~B
  \textbf{60}, 2921(1999).

\bibitem{niemi1} J. Nieminen, E. Niemi, V. Simic-Milosevic, and
  K. Morgenstern, Phys. Rev. B \textbf{72}, 195421 (2005).

\bibitem{Moon2008}C. R. Moon, L. S. Mattos, B. K. Foster, G. Zeltzer, W.
  Ko and H. C. Manoharan, Science \textbf{319}, 782(2008).

\bibitem{teha} J. Tersoff, and D. R. Hamann, Phys. Rev. B \textbf{31},
  805 (1985).

\bibitem{todorov} T. Todorov, G. Briggs, and A. Sutton, J.  Phys.:
  Cond. Matt. \textbf{5}, 2389 (1993).

\bibitem{pendry} J. B. Pendry, A. Pr\^etre, and B. Krutzen, J.  Phys.:
  Cond. Matt. \textbf{3}, 4313 (1991).

\bibitem{VASP} G. Kresse and J. Furthm\"uller, Phys. Rev. B
  \textbf{54}, 11169 (1996).

\bibitem{slat} J. C. Slater and G. F. Koster, Phys. Rev.  \textbf{94},
  1498 (1954).

\bibitem{Horsfield} A. P. Horsfield, A. M. Bratkovsky, D. G. Pettifor,
  and M. Aoki, Phys. Rev. B \textbf{53}, 1656 (1995).

\bibitem{Harrison} W.A. Harrison, Electronic Structure and the
  properties of solids - The Physics of The Chemical Bond, (Dover
  publications, INC., New York, 1989).

\bibitem{niemi2} E. Niemi, V. Simic-Milosevic, K. Morgenstern,
  A. Korventausta, S. Paavilainen, and J. Nieminen,
  J. Chem. Phys. \textbf{125}, 184708 (2006).

\bibitem{akorunpub} A. Korventausta, S. Paavilainen, J. Nieminen,
  unpublished.

\bibitem{spinendnote}The SOMO state dominating the STM imaging is
  splitted in the spin-polarized DFT calculations showing a gap of
  about 1.0 eV. This does not, however, change the shape of the
  orbitals, and thus the STM images. The bias voltage used in
  spin-polarized calculations has to be large enough to include the
  SOMO state while in non-spin-polarized calculations a small value is
  adequate.

\bibitem{hoffmann} J. Hoffmann, J. Chem. Phys.  \textbf{39}, 1397
  (1963).

\bibitem{anderson} A. B. Anderson, J. Chem. Phys.  \textbf{62}, 1187
  (1975)

\bibitem{mulli49} R. S. Mulliken, C. A. Rieke, D. Orloff, and
  H. Orloff, J. Chem. Phys. \textbf{17}, 1248 (1949).

\bibitem{clera} E. Clementi and D. L. Raimondi,
  J. Chem. Phys. \textbf{38}, 2986 (1963).

\bibitem{paavi2} J. Repp, G. Meyer, S. Paavilainen, F. Olsson and
  M. Persson, Phys. Rev. Lett. \textbf{95}, 225503 (2005).

\bibitem{MW} Y. Meir and N.S. Wingreen, Phys. Rev. Lett. \textbf{68},
  2512 (1992).

\bibitem{NessFisher} H. Ness and A.J. Fisher, Phys. Rev. B \textbf{56}, 12469 (1997).

\bibitem{McKinnon} B.A. McKinnon and T.C.Choy, Phys. Rev. B \textbf{54}, 11777 (1996).

\bibitem{Mingo} N.Mingo, L. Jurczyszyn, F.J. Garcia-Vidal,
  R. Saiz-Pardo, P.L. de Andres, F. Flores, S.Y. Wu, and W. More,
  Phys. Rev. B \textbf{54}, 2225 (1996).

\end{thebibliography}

\end{document}